\documentclass[twocolumn,preprintnumbers,amsmath,amssymb,pre]{revtex4}

\usepackage[]{graphicx}

\usepackage{dcolumn}
\usepackage{bm}

\begin{document}

\title{Universal velocity distributions in an experimental granular fluid}


\author {P.M. Reis\footnote{Current address: Laboratoire PMMH (UMR7636 CNRS-ESPCI-P6-P7), 10 rue Vauquelin, 75231 Paris, FRANCE. \texttt{preis@pmmh.espci.fr} }, R.A. Ingale and M.D. Shattuck}
\affiliation{ Benjamin Levich Institute, The City College of the
City University of New York \\ 140th St. and Convent Av., New York
NY 10031, USA }

\begin{abstract}

We present experimental results on the velocity statistics of a
uniformly heated granular fluid, in a quasi-2D configuration. We find
the base state, as measured by the single particle velocity
distribution $P(c)$, to be universal over a wide range of filling
fractions and only weakly dependent on all other system
parameters. There is a consistent overpopulation in the distribution's
tails, which scale as $ P\propto\exp(\mathrm{-A}\times
c^{-3/2})$. More generally, $P(c)$ deviates from a Maxwell-Boltzmann
by a second order Sonine polynomial with a single adjustable
parameter, in agreement with recent theoretical analysis of inelastic
hard spheres driven by a stochastic thermostat. To our knowledge, this
is the first time that Sonine deviations have been measured in an
experimental system.

\end{abstract}

\maketitle


\section{Introduction}
\label{sec:introduction}

The study of granular flows has received a recent upsurge of interest
in physics\cite{jaeger:1996}. This has been motivated by both the
relevance of such flows to a wide range of industrial and geological
processes, and by the realization that granular materials provide an
excellent test bed for a number of fundamental question in the context
of modern fluid dynamics and non-equilibrium statistical mechanics
\cite{kadanoff:99}.  Granular media are ensembles of macroscopic
particles, in which kinetic energy ($K_E$) can be preferentially injected
in the $N\simeq 10-10^{10}$ translational and rotational degrees of
freedom of the center of mass of the particles without effecting the
$N_A\simeq10^{23}$ internal degrees of freedom of each particle.
Therefore the {\em granular} temperature $T=K_E/N$ is $10-20$ orders of
magnitude larger than the molecular temperature $\tau=K_E/N_A$. In this
sense the molecular temperature is irrelevent.  Moreover, granular
materials are intrinsically dissipative since energy from the $K_E$ of
the particles is lost (converted) through inelastic collisions and
frictional contacts to the molecular temperature. Hence, any dynamical
study of a granular media requires an energy input, which typically
takes the form of vibration or shear\cite{melo:1994,miller:1996}. This
interplay between energy injection and dissipation forces the
system to be far from equilibrium and granular fluids are therefore
not expected to behave similarly to the equilibrium counter-parts of
classical fluids\cite{kadanoff:99}.

However, we have recently shown that the structure of a uniformly
heated quasi-2D granular fluid is identical to that found in
simulations of equilibrium hard disks\cite{reis:2006}.  Further, we
find that the mean-squared displacement of grains shows diffusive
behavior at low density and caging behavior as density is increased
until particles are completely confined above freezing point just as
in ordinary molecular fluids\cite{reis:caging:2006}.  These surprising
results provide a partial experimental justification for the
theoretical treatment of granular flow by analogy with molecular
fluids using standard statistical mechanics and kinetic theory.  A
large number of kinetic theories of granular flows{\cite{savage:1981,
lun:1983, jenkins:1985, jenkins:85, jenkins88, lun91, goldshtein95,
sela96, brey96, santos98, vannoije98b, garzo99, dufty03, brey04,
dufty04, kumaran05, brey05, lutsko06}} have been produced starting in
1980's.  The results of these theories are balance equation analogous
to the Navier-Stokes equations for the density, $n$, the average
velocity, $\mathbf{v}_o$, and the average kinetic energy per particle
or temperature, $T$, as well as, equations of state and transport
coefficients, which are functions of $n$,$\mathbf{v}_0$, and $T$.  The
heart of these formulations is the single particle distribution
function $f^{(1)}\left( \mathbf{r},\mathbf{v},t\right)$, which
measures the probability of a particle a position $\mathbf{r}$ having
velocity $\mathbf{v}$.  The field equations are derived from moments
of $f^{(1)}\left( \mathbf{r},\mathbf{v},t\right)$.  For example, the
density, $n$, is the 0th moment, the average velocity, $\mathbf{v}_o$,
is the 1st moment, and the temperature, $T$, is the 2nd moment.  The
transport coefficients are deterimined from non-equilibrium
perturbations of $f^{(1)}\left( \mathbf{r},\mathbf{v},t\right)$.
Generally the equilibrium or steady-state distribution is assumed to
be indepent of space and to have Maxwell-Boltzmann form
$f^{(1)}=f_{MB}=A\exp \left(
-\frac{|\mathbf{v}-\mathbf{v}_0|^2}{2T}\right)$, although, recent work
has explored the use of other base states or steady-state distributions
\cite{dufty04,brey05,lutsko06}.

Two important questions one may ask are: 1) What is the base state
$f^{(1)}$ for a driven granular fluid? and 2) Does it have a universal
form?. If it is characterized by a small number of parameters such as
the moments of the $f^{(1)}$ distribution as in the case for ideal
fluids, one would be able to advance a great deal in developing
predictive continuum models for granular flows since the theoretical
machinery from kinetic theory could be readily applied.  The careful
experimental study and characterization of this base state is
therefore important in order to develop a theory for granular media
similar to that of regular molecular liquids.

There has been a substantial amount of experimental, numerical, and
theoretical work to address these issues in configurations where the
energy input perfectly balances the dissipation such that the system
reaches a steady state, albeit far from equilibrium. These
\emph{non-equilibrium steady states} are simplified realizations of
granular flows more amenable to analysis, but the insight gained from
their study should be helpful in the tackling of other more complex
scenarios. One feature that has been consistantly established in
experiments is that single particle velocity distribution functions
deviate from the Maxwell-Boltzmann distribution
\cite{losert:1999,olafsen:1999,rouyer:2000,blair:2001,prevost:2002,aranson:2002,blair:2003}. In
particular, the tails (i.e. large velocities) of the experimental
distributions exhibit a considerable overpopulation and have been
shown to scale as $f^{(1)}(v_i)\sim\exp [(v_i / \sqrt{T_i})^{3/2}]$
where $v_i$ is a velocity component and $T_i=\langle v_i^2\rangle$ is
the \emph{granular temperature}. This behavior is in good agreement
with both numerical \cite{moon:2001} and theoretical
\cite{vannoije:1998} predictions. Even though the tails correspond to
events with extremely low probabilities, they increase the variance of
the distribution.  The variance of the distribution is $T$, the
average kinetic energy of the particles.  Using a Gaussian to
represent this type of distribution leads to major discrepancy in the
region of high probability at the central part of the distribution
(low velocities) \cite{blair:2003} which have, thus far, been greatly
overlooked in experimental work.  Note that requiring the variance
of $f^{(1)}(v_i)$ to be the granular temperature greatly
constrains the functional form that the distribution can take.

A model system that has been introduced to study this question is the
case of a homogeneous granular gas heated by a stochastic thermostat,
i.e. an ensemble of inelastic particles randomly driven by a white
noise energy source \cite{williams:96}. Recently, there have been many
theoretical and numerical studies on this model system
\cite{vannoije:1998,vannoije:1999,montanero:2000,moon:2001,brilliantov:2006,poschel:2006}
where the steady state velocity distribution have been found to
deviate from the Maxwell-Boltzmann distribution. van Noije and Ernst
\cite{vannoije:1998} studied these velocity distributions based on
approximate solutions to the inelastic hard sphere Enskog-Boltzmann
equation by an expansion in Sonine polynomials. The results of their
theoretical analysis has been validated by numerical studies using
both molecular dynamics \cite{moon:2001} simulations and direct
simulation Monte Carlo \cite{montanero:2000,poschel:2006}.  The use
Sonine corrections is particular attractive since it leaves the
variance of the resulting velocity distribution unchanged but leads to
a non-Gaussian fourth moment or kurtosis $K\ne 3$.

We have addressed the above issues in a well controlled experiment in
which we are able to perform precision measurements of the velocity
distributions of a uniformly heated granular fluid. A novel feature of
our experimental technique is that we are able to generate macroscopic
random walkers over a wide range of filling fractions and thermalize
the granular particles in a way analogous to a stochastic thermostat.
This is an ideal system to test the applicability of some of the
kinetic theory ideas mentioned above. In our earlier study
\cite{reis:2006}, we focused on the \emph{structural configurations}
of this granular fluid and revealed striking similarity to those
adopted by a fluid of equilibrium hard disks.  Here we concentrate on
the \emph{dynamical aspects} of this same experimental system, as
measured by the single particle velocity distribution, $P(c)$, and
observe a marked departure from the equilibrium behavior, i.e. from
the Maxwell-Botlzmann distribution. We quantify these deviations from
equilibrium and show that they closely follow the predictions, in the
kinetic theory framework, of the theoretical and numerical work
mentioned above. In particular, we find a consistent overpopulation in
the distribution's high energy tails, which scale as stretched
exponentials with exponent $-3/2$. In addition, we experimentally
determine the deviations from a Gaussian of the full distributions
using a Sonine expansion method and find them to be well described by
a second order Sonine polynomial correction. We establish that the
{\em entire} experimental velocity distribution functions are
\emph{universal} and well described by
	\begin{equation}
	P(c)=f_{MB}\left(1+a_{2}(1/2c^{4}-3/2^{2}+3/8)\right)
	\end{equation}
where $a_{2}$ is the non-gaussianity parameter that can be related to the Kurtosis of the distribution and the forth-order polynomial is the Sonine polynomial of order 2. This functional form of $P(c)$ was found to be valid over a wide range of the system parameters. To our knowledge, this is the first time that the Sonine corrections  of the velocity distributions are measured in an experimental system, in agreement with analytical predictions.
 
This paper is organized as follows. In Sec.
\ref{sec:kinetic_theory} we briefly review the theoretical framework
of kinetic theory along with the steady state solution to the
Enskog-Boltzmann equation using the Sonine polynomial expansion
method. In Sec. \ref{sec:apparatus} we present our experimental
apparatus and describe how the homogeneous granular gas with random
heating is generated. In Sec. \ref{sec:singleparticledriving} we
study the thermalization mechanism by analyzing the trajectories of a
single particle in the granular cell. In Sec.,
\ref{sec:granaulartemperature}, using the concept of granular
temperature, we then quantify the dynamics of the experimentally
obtained non-equilibrium steady states, for a number of parameters,
namely: the filling fraction, the driving parameters (frequency and
acceleration) and the vertical gap of the cell. In Sec.
\ref{sec:pofv} we turn to a detailed investigation of the Probability
Distribution Functions of velocities as a function of the system
parameters. In particular, we quantify the deviations from Maxwell-Boltzmann behavior at large velocities (tails of the distributions) -- Sec. \ref{sec:pofv_tails}. In Sec. \ref{sec:pofv_lowv} we extend the deviation analysis to the full range of the distribution using an expansion method that highlights deviations at low velocities and allows us to make a connection with Sonine polynomials. We finish in Sec. \ref{sec:conclusion} with some
concluding remarks.


\section{Brief Review of Theory}
\label{sec:kinetic_theory}

We briefly review the key features of the kinetic theory for granular
media pertinent to our study.  The number of particles in a volume
element, $\textsl{d}\textbf{r}$, and velocity element,
$\textsl{d}\textbf{v}$, centered at position $\textbf{r}$ and velocity
$\textbf{v}$ is given by
$f^{1}(\textbf{r},\textbf{v},t)\textsl{d}\textbf{r}\textsl{d}\textbf{v}$,
where $f^{1}(\textbf{r},\textbf{v},t)$ is the single particle
distribution function. Continuum fields are given as averages over
$f^{1}(\textbf{r},\textbf{v},t)$. For instance, the number density,
$n$, average velocity, $v_o$, and granular temperature, $T$, are given
respectively by,
\begin{equation}
n(\textbf{r},t)\equiv \int
f^{1}(\textbf{r},\textbf{v},t)\textsl{d}\textbf{v},
\end{equation}
\begin{equation}
\textbf{v}_o(\textbf{r},t)\equiv \frac{1}{n}\int
f^{1}(\textbf{r},\textbf{v},t)\textbf{v}\textsl{d}\textbf{v},
\end{equation}
\begin{equation}
T(\textbf{r},t)\equiv \frac{1}{nd}\int
f^{1}(\textbf{r},\textbf{v},t)(\textbf{v}-\textbf{v}_{o})^{2}\textsl{d}\textbf{v},
\end{equation}
where $d$ is the number of dimensions. It is important to note
that the granular temperature $T$ is not the thermodynamic
temperature but, by analogy, the kinetic energy per macroscopic particles (explained in detail in Sec.
\ref{sec:granaulartemperature}). For the spatially homogeneous and isotropic
case, we refer to the single particle distribution function as
$f^{1}(\textbf{v},t)$ and consider only a single component of the velocity, i.e. $f^{(1)}(v_i,t)$.
From now on we drop the index $i$. It is also convenient to introduce a
scaled distribution function $f^{1}(c,t)$ where the velocity is scaled
by a characteristic velocity such that, $c=v/\sqrt{2T(t)}$, where
$T(t)$ is the granular temperature and equal to the variance of $f^{1}(c,t)$.

The stocastically heated single particle distribution function
$f(c,t)$ satisfies the \emph{Enskog-Boltzmann} equation,
    \begin{equation}
        \frac{\partial f}{\partial
        t}=\mathcal{Q}(f,f)+\mathcal{F}_{FP}f,
        \label{eqn:enskog_boltzmann}
    \end{equation}
where, $\mathcal{Q}$, is the collision operator, which accounts for
the inelastic particle interactions and the Fokker-Plank operator,
$\mathcal{F}_{FP}$, accounts for the stochastic forcing
\cite{vannoije:1998,vannoije:98}.  We are interested in a
stationary solution of Eqn. (\ref{eqn:enskog_boltzmann}) where the
heating exactly balances the loss of energy due to collisions and
the temperature becomes time independent.  van Noije and Ernst
\cite{vannoije:1998} obtained steady state solutions to Eqn.
(\ref{eqn:enskog_boltzmann}) by taking the series expansion of
$f(c)$ away from a Maxwell-Boltzmann, $f_{MB}$, i.e.
    \begin{equation}
        f(c)=f_{MB}(c) \left\{ 1+\sum_{p=1}^\infty a_p S_p(c^2)\right\},
        \label{eqn:sonine_expansion}
    \end{equation}
in terms of the Sonine polynomials $S_p(c^2)$ where $a_p$ is a
numerical coefficient for the $p^{th}$-order. The deviations from
the MB distribution are thus expressed in terms of an expansion on Sonine
polynomials. The Sonine polynomials, which are knows as
associated Laguerre polynomials in different contexts, are defined
as,
\begin{equation}
S_p(c^2)=\sum_{n=0}^p \frac{(-1)^n (p+1/2)!}{(n+1/2)! (p-n)!n!}
(c^2)^n. \label{eqn:sonine_poly}
\end{equation}
These are orthogonal accordingly to the  relationship,
\begin{equation}
\int_{0}^{\infty} c^2 e^{-c^2} S_p(c^2)S_m(c^2) \textit{dc}
 =\frac{1}{2} \delta_{pm} \frac{(\frac{1}{2}+n)!}{n!}.
\end{equation}
To further simply the calculation, terms higher than
$\mathcal{O}(2)$ are typically neglected, such that,
    \begin{equation}
        f(c)=f_{MB}(c) \left\{ 1+ a_2S_2(c^2)\right\},
        \label{eqn:theory_truncation}
    \end{equation}
where,
\begin{equation}
f_{MB}(c)=\frac{1}{\sqrt{\pi}}exp(-c^2),
\label{eqn:MB}
\end{equation}
is the Maxwell-Boltzmann distribution and
\begin{equation}
S_2(c^2)=\frac{1}{2}c^4-\frac{3}{2}c^2+\frac{3}{8}
\end{equation}
is the $\mathcal{O}(2)$ Sonine polynomial for 2D. The first Sonine
coefficient, $a_1$, is zero according to the definition of
temperature \cite{goldshtein:1995}. The second Sonine coefficient,
$a_2$ is the first non-trivial Sonine coefficient and hence the
first non-vanishing correction to the MB distribution and can be
related to the Kurtosis $K=3(a_2+1)$ of the distribution.


\section{Our experiments}
    \label{sec:apparatus}

\begin{figure}[b]
    \begin{center}
            \includegraphics[width=7cm]{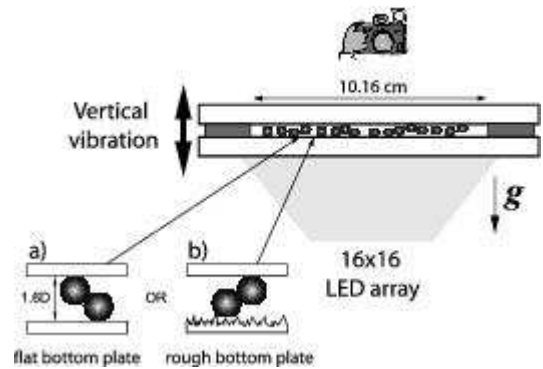}
        \caption{Schematic diagram of the experimental apparatus. The ensemble of stainless steel spheres was sandwiched in between two glass plates, separated by a $1.905mm$ thick annulus (dark grey). Top glass plate was always optically flat. Bottom glass plate could either be (a) optically flat or (b) rough by sand-blasting.
        \label{fig:apparatus}}
    \end{center}
\end{figure}

In Fig.~\ref{fig:apparatus} we present a schematic diagram of our
experimental apparatus, which is adapted from a design introduced by
Olafson and Urbach \cite{olafsen:1999,prevost:2002}. An ensemble of
stainless steel spheres (diameter $D=1.19mm$) were vertically vibrated
in a shallow cylindrical cell, at constant volume conditions. The cell
consisted of two parallel glass plates set horizontally and separated
by a stainless steel annulus. This annulus had an inner diameter of
$101.6mm$ and, unless otherwise stated, was $1.905mm$ thick, which set
the height of the cell to be 1.6 particle diameters. This constrained
the system to be quasi-two dimensional such that each sphere could not
fully overlap.  However the 2D projection of the particles have a
minimum separation of $r_{min}=0.8D$ (rather than $r_{min}Õ=D$ for a
strictly two-dimensional system). Hence, the extreme situation of a
maximum overlap between two particles ($\delta=D-r_{min}=0.2D$)
occurred when one of the particles of the adjacent pair was in contact
with the bottom plate and the other in contact with the top plate, as
shown in Fig.  \ref{fig:apparatus}(a). We have also performed
experiments to explore the effect of changing the height of the cell.

The top disk of the cell was made out of an optically flat borofloat
glass treated with a conducting coating of ITO to eliminate
electrostatic effects. For the bottom disk of the cell, two variants
were considered: a flat glass plate identical to the top disk --
Fig.~\ref{fig:apparatus}(a) -- and a rough borosilicate glass plate --
Fig.~\ref{fig:apparatus}(b). In the second case, the plate was
roughened by sand-blasting, generating random structures with
lengthscales in the range of $50$--$500 \mu m$. The use of a rough
bottom glass plate improves on the setup of Olafsen and Urbach
\cite{olafsen:1999}, who used a flat plate. As it will be shown in
Sec.  \ref{sec:singleparticledriving} and Appendix \ref{sec:appendix},
it had the considerable advantage of effectively randomizing the
trajectories of single particles under vibration, allowing a
considerably wider range of filling fractions to be explored.

The horizontal experimental cell was vertically vibrated,
sinusoidally, via an electromagnetic shaker (VG100-6 Vibration
Test System). The connection of the shaker to the cell was done
via a robust rectangular linear air-bearing which constrained the
motion to be  unidirectional. The air-bearing
ensured filtering of undesirable harmonics and non-vertical
vibrations due to its high stiffness provided by high-pressure air
flow in between the bearing surfaces. Moreover, the coupling
between the air-bearing and the shaker consisted of a thin brass
rod ($25.4mm$ long and $1.6mm$ diameter). This rod could slightly
flex to correct any misalignment present in the shaker/bearing
system, while being sufficiently rigid in the vertical direction
to fully transmit the motion. This ensemble -- shaker/brass
rod/air-bearing -- ensured a high precision vertical oscillatory
driving.

The {\em forcing}
parameters of the system were the frequency, $f$, and amplitude,
$A$, of the sinusoidal oscillations. From these, it is common
practice to construct a non-dimensional acceleration parameter,
     \begin{equation}
               \Gamma=\frac{A(2\pi f)^2}{g},
          \label{eqn:Gamma}
     \end{equation}
where $g$ is the gravitational acceleration. We worked within the
experimental ranges $(10<f<100)Hz$ and $1<\Gamma <6$. The first {\em
control} parameter was the filling fraction of the steel spheres in
the cell defined as,
     \begin{equation}
               \phi=\frac{N\pi (D/2)^2}{\pi R^2},
          \label{eqn:phi}
     \end{equation}
where $N$ is the number of spheres (with diameter $D=1.19mm$) in the
cell of radius $R=50.8mm$. The filling fraction $\phi$ is therefore
defined in the projection of the cell onto the horizontal plane. The
second control parameter was the height, $h$, of the experimental cell
which we varied from $1.3D<h<2.3D$.

The granular cell was set horizontal in order to minimize compaction
effects, inhomogeneities and density gradients which otherwise would
be induced by gravity. This way, a wide range of filling fractions,
$0<\phi\leq 0.8$, could be accurately explored by varying the number
of spheres in the cell, $N$, down to a resolution of single particle
increments. Moreover, as we shall show in
Sec. \ref{sec:singleparticledriving}--\ref{sec:granaulartemperature},
we were able to attain spatially uniform driving of the spheres in the
cell due to the use of the rough glass bottom plate.

The dynamics of the system was imaged from above by digital
photography using a grayscale DALSA CA-D6 fast camera, at 840
frames per second. The granular layer was illuminated from below,
in a transmission configuration, by a $16\times 16$ array of  high
intensity LEDs. In this arrangement the particles obstruct the 
light source and appear as dark circles in a bright background.

We have developed particle tracking software based on a
two-dimensional least squares minimization of the individual
particle positions, which is able to resolve position to sub-pixel
accuracy. By focusing on a $(15\times15)mm^2$ imaging window
located in the central region of the full cell we were typically
able to achieve resolutions of 1/20--1/10 of a pixel (which
corresponds to $2.5$--$5\mu m$). From the trajectories of the
particles we could easily calculate an approximately instantaneous
discretized velocity as,
     \begin{equation}
          |v^i(t_j+\Delta t)|= \sqrt{(v_x(t_j+\Delta t))^2 + v_y((t_j+\Delta t))^2}
          \label{eqn:velocitydefinition}
     \end{equation}
with
     \begin{equation}
     v_x=\frac{x^i(t_j+\Delta t)-x^i(t)}{\Delta t} \quad
     \mathrm{and} \quad
     v_y=\frac{y^i(t_j+\Delta t)-y^i(t)}{\Delta t}
     \end{equation}
for the $i^{th}$ particle in an experimental frame at time $t_j$
where $\Delta t=1.19\mu s$ is the time interval between two frames
and $(x,y)$ are the cartesian coordinates on the horizontal plane.
For the remainder of this paper we shall be interested in the
nature of the probability distribution function of velocities,
$P(v)$, of the driven particles. Statistics for calculating the
$P(v)$ distributions were constructed by analyzing 2048 frames at
an acquisition rate of 840 frames per second which corresponded to
an acquisition real time of $2.4381s$.


\section{Single particle driving}
\label{sec:singleparticledriving}

\begin{figure}[b]
    \begin{center}
            \includegraphics[width=6cm]{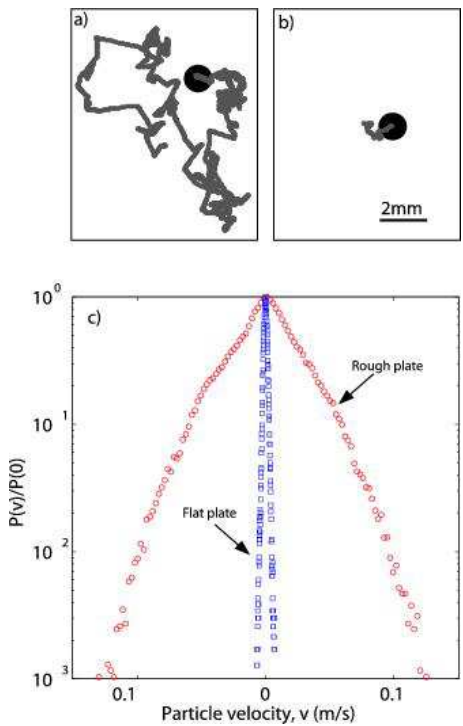}
        \caption{Typical trajectories for single particle driven using a (a) rough or (b) flat bottom glass plate. Each of the trajectories is $2.438s$ long. (c) Probability distribution function of velocities for a single particles using rough ($\circ$) and flat ({$\square$}) bottom glass plate. Driving parameters: $f=50Hz$, $\Gamma=4.0$. \label{fig:pofv_singleparticles_pofv}}
    \end{center}
\end{figure}

We first concentrate on the case of the driving of a single
particle in the vibrating experimental cell.  For each
experimental run, a single sphere was positioned in the field of
view of the camera with the aid of a strong magnet temporarily
placed beneath the imaging window. The magnet was then removed and
the trajectory of the sphere immediately recorded. This was
repeated 100 times to obtain statistics. Typical trajectories for
two runs with a flat and rough bottom glass plates are shown in
Fig. \ref{fig:pofv_singleparticles_pofv}(a) and (b), respectively.
In both cases, the cell was vibrated at a frequency of $f=50Hz$
and dimensionless acceleration of $\Gamma=4.0$.  During each cycle
of the driving, the motion of the single particle was thermalized
due to collisions with both top and bottom glass plates.

By comparing Fig. \ref{fig:pofv_singleparticles_pofv}(a) with Fig.
\ref{fig:pofv_singleparticles_pofv}(b), it is clear that the
particle excursions obtained by using a rough bottom plate are
considerable larger than those for a flat plate. This can be
readily quantified by calculating the Probability Density
Functions of the velocities of single particles, $P(v)$, which we
plot in Fig. \ref{fig:pofv_singleparticles_pofv}(c), for both
cases. We now define the average kinetic energy, also known as
\emph{granular temperature}, of a single particle as the variance
of its velocity distribution,
     \begin{equation}
          T=T_x+T_y=\frac{1}{2}\left(\langle v_x^2\rangle + \langle                                                    v_y^2\rangle \right),
          \label{eqn:singleparticleT}
     \end{equation}
where $v_x$ and $v_y$ are the two orthogonal components of the
velocity in the 2D horizontal plane and the brackets $\langle .
\rangle$ denote time averages for the timeseries of the velocity
components for 100 particle trajectories. Note that since we are
dealing with monodisperse spheres, in this definition of kinetic
temperature, the mass of the particle is taken to be unity. We
have measured $T^s_{rough}=9.53\times 10^{-4} m^2s^{-2}$ if a
rough bottom glass plate is used  and $T^s_{flat}=2.51\times
10^{-6} m^2s^{-2}$ for the case of the flat bottom glass plate.
This yields a temperature ratio between the two cases of
$T^s_{rough}/T^s_{flat}=380$.

Indeed, this is a signature that  the structures in the rough
plate were considerably more effective than the flat plate in
transferring and randomizing the momentum of the steel spheres in
the vertical direction (due to the sinusoidal vertical oscillation
of the cell) into the horizontal plane. The velocity of the steel
sphere was randomized each time it collided with the peaks and
valleys randomly distributed across the the
sand-blasted rough glass surface.  We can therefore regard this
system as a \emph{spatially uniform heater} of the granular
particles, thereby generating macroscopic random walkers. For this
reason, and as it will become more clear in the next Sec.,
where we will further comment on the issue of isotropy, we shall
focus our experimental study in using the rough bottom glass
plate.


\section{Driven monolayers: granular temperature}
\label{sec:granaulartemperature}

\begin{figure}[b]
    \begin{center}
        \begin{tabular}{lll}
            a) & b) & c) \\
            \includegraphics[width=2.7cm]{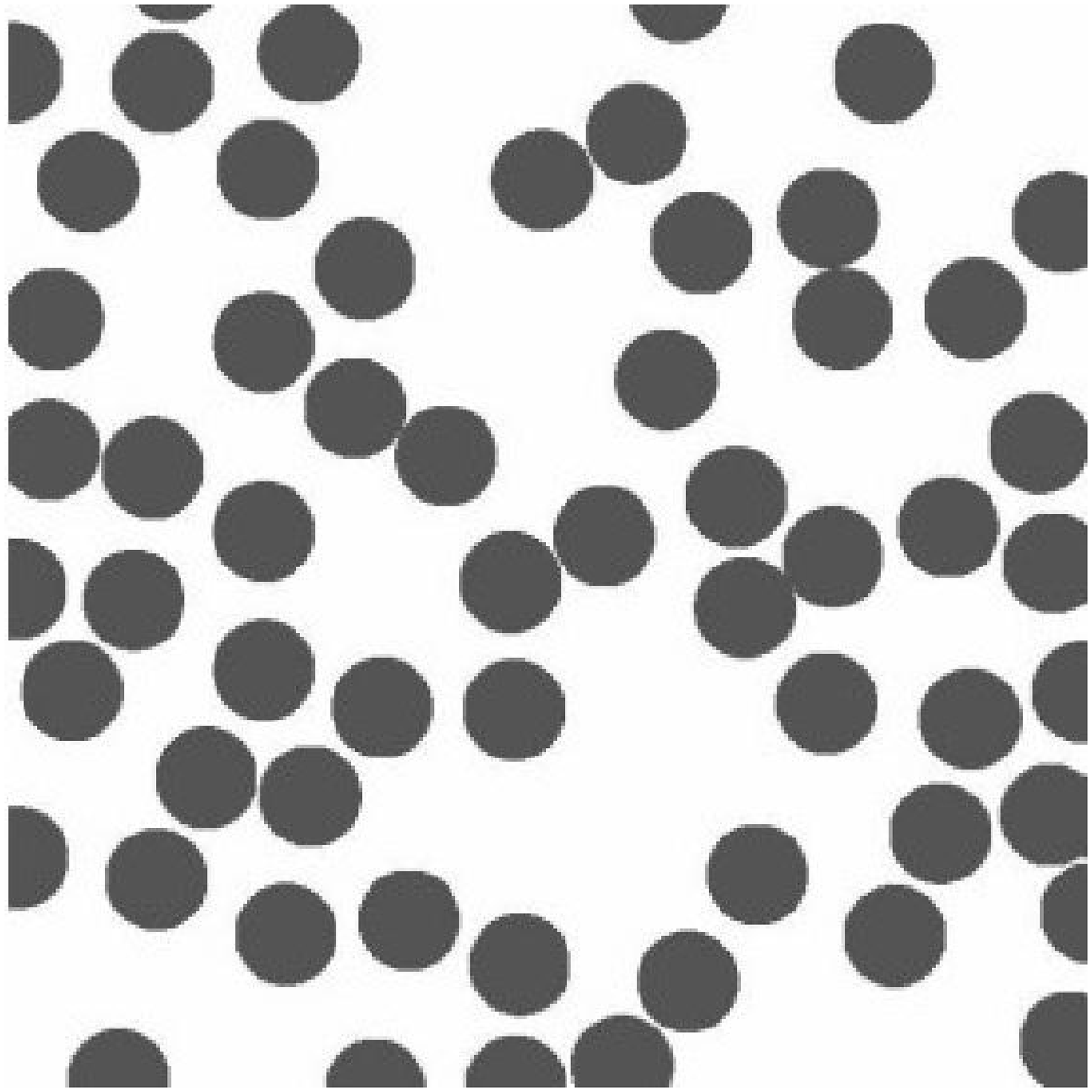} &
            \includegraphics[width=2.7cm]{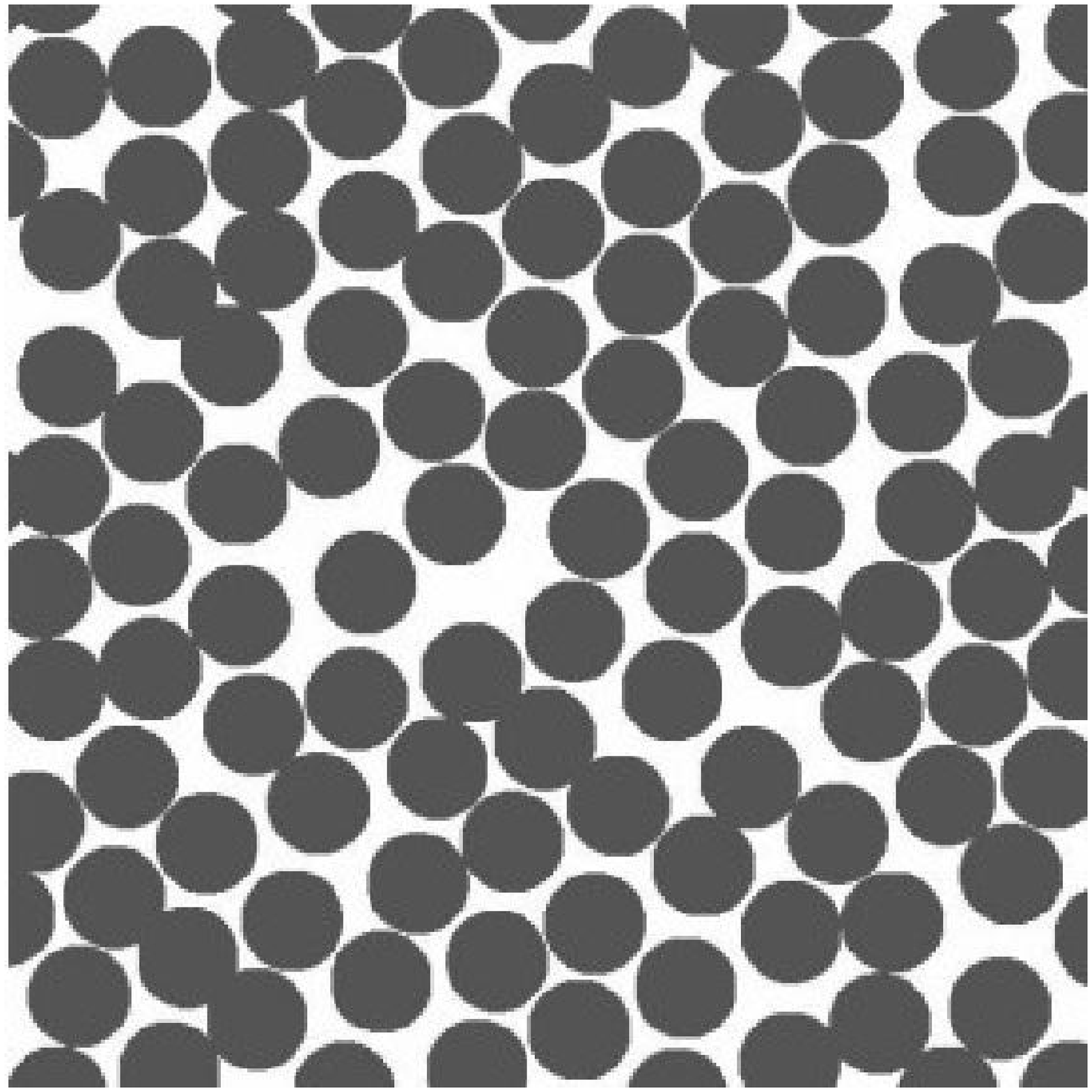} &
            \includegraphics[width=2.7cm]{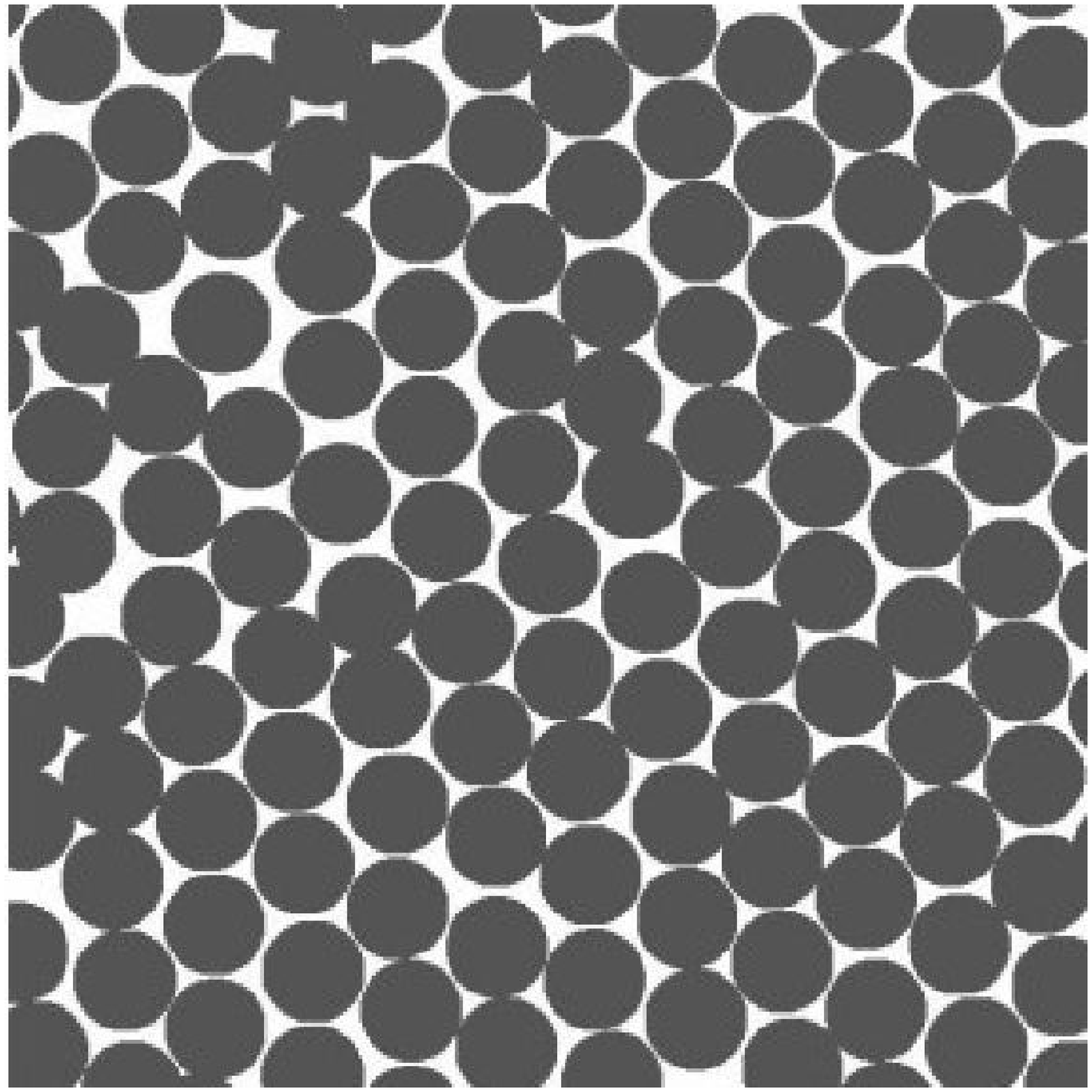}
        \end{tabular}
        \caption{Snapshots of typical configurations of the granular layer at various values of the filling fraction: (a) dilute gas, $\phi=0.34$, (b) dense liquid, $\phi=0.67$ and (c) thermalized crystal, $\phi=0.80$. Driving parameters: $f=50Hz$, $\Gamma=4.0$. \label{fig:experimentalframes}}
    \end{center}
\end{figure}

Having investigated the dynamics of single particles in the granular
cell, we now turn to the study of granular monolayers at higher
filling fractions. In Fig. \ref{fig:experimentalframes} we present
typical configurations, at three values of $\phi$, for a granular
layer driven at $f=50Hz$ and $\Gamma=4$ with a rough bottom plate. The
snap-shot in Fig.  \ref{fig:experimentalframes}(a), for $\phi=0.34$,
corresponds to a dilute state in which the particles perform large
excursions in between collisions as they randomly diffuse across the
cell. If the filling fraction is increased, as in the example of the
frame shown in Fig. \ref{fig:experimentalframes}(b) for $\phi=0.67$,
one observes a higher collision rate characteristic of a dense gaseous
regime. For even higher values of filling fraction the spheres ordered
into an hexagonally packed arrangement and became locked into the
\emph{cage} formed by its six neighbors. The system is then said to be
\emph{crystalized} as shown in the typical frame presented in
Fig. \ref{fig:experimentalframes}(c) for $\phi=0.80$. The structural
configurations associated with the crystallization transition, as a
function of filling fraction, were studied in detail in
\cite{reis:2006} and the \emph{caging} dynamics, as crystallization is
approached in \cite{reis:caging:2006}.

As  defined in the previous section the granular
temperature is the average kinetic energy per particle. For this
the brackets $\langle . \rangle$ in Eqn.
(\ref{eqn:singleparticleT}) now denote both time and ensemble
averages for all the spheres found within the imaging window.
Moreover, one can define $T_x=1/2\langle v_x^2\rangle$ and
$T_y=1/2\langle v_y^2\rangle$, with $T=T_x+T_y$, as the temperature
projections onto the $x$ and $y$ directions of the horizontal
plane.

\begin{figure}[t]
    \begin{center}
            \includegraphics[width=8cm]{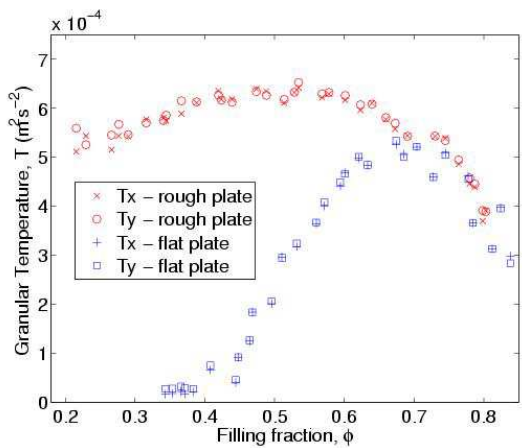}
        \caption{Filling fraction dependence of the granular temperature
        ($T_{x}$ and $T_{y}$ are the $x$ and $y$ components, respectively),
        for both cases of using a flat and rough bottom glass plates.
        Driving parameters: $f=50Hz$ and $\Gamma=4.0$. \label{fig:temperatures_flat_rough}}
    \end{center}
\end{figure}

In Fig. \ref{fig:temperatures_flat_rough} the temperature components,
$T_x$ and $T_y$, are plotted as a function of filling fraction. For
both cases of using a rough and a flat bottom plate, the respective
$T_x$ and $T_y$ are identical. This shows that the forcing of the
granular particles was isotropic. In agreement with the case of a
single particle, the dynamics of the granular layer using a flat or a
rough bottom plate is remarkably different. If the rough bottom plate
is used, the granular temperature depends almost monotonically on the
filling fraction. At low $\phi$, $T$ is approximately constant as the
layer simply feels the Ô\emph{thermal bath}Õ and shows little increase
in $T$ until $\phi\sim0.5$ is reached.  As the filling fraction is
increased past $\phi \sim 0.5$, the granular temperature rapidly
decreases due to energy loss in the increasing number of collisions
and due to a decreasing available volume.

\begin{figure}[b]
    \begin{center}
      \includegraphics[width=6cm]{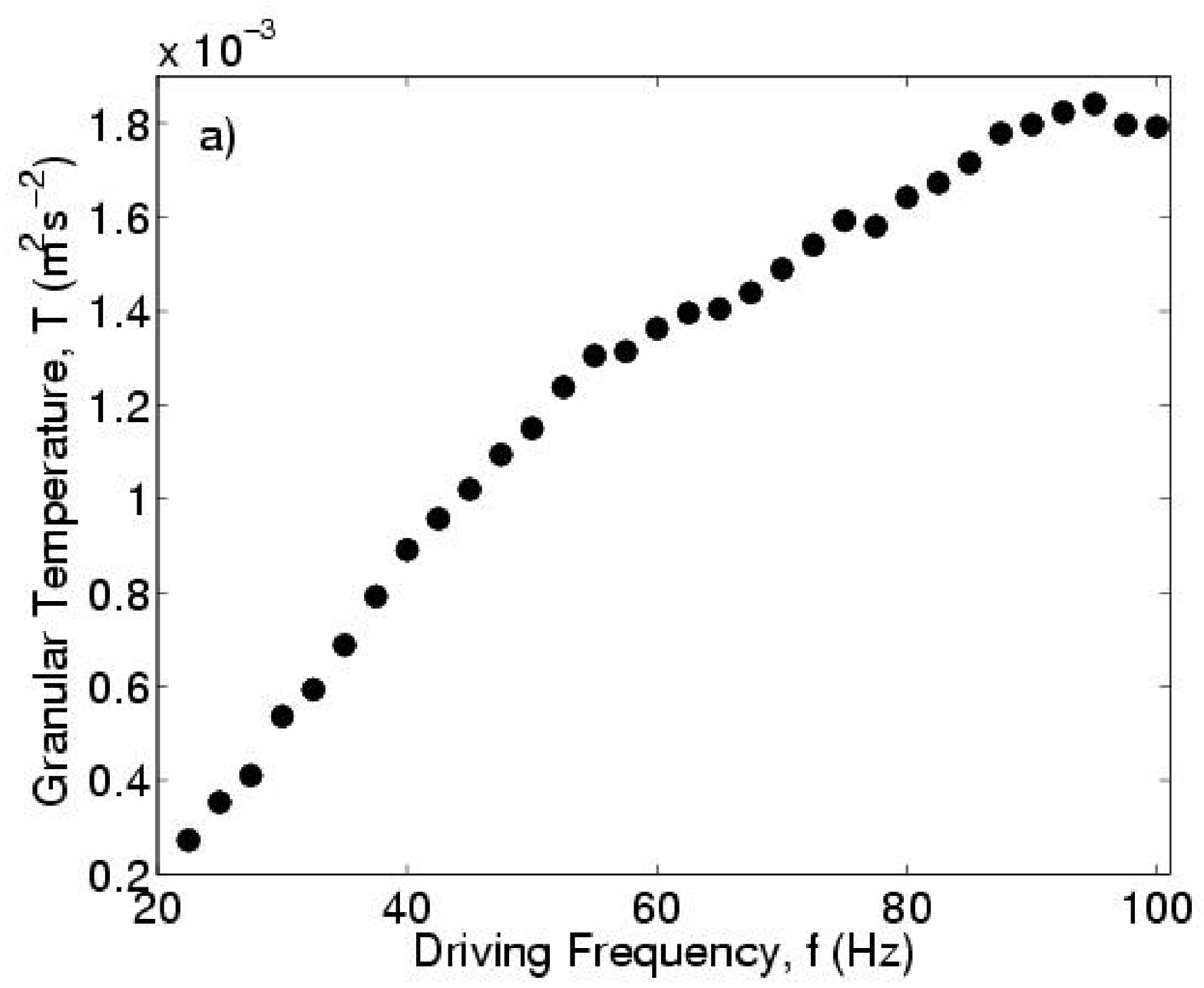}\\
      \includegraphics[width=6cm]{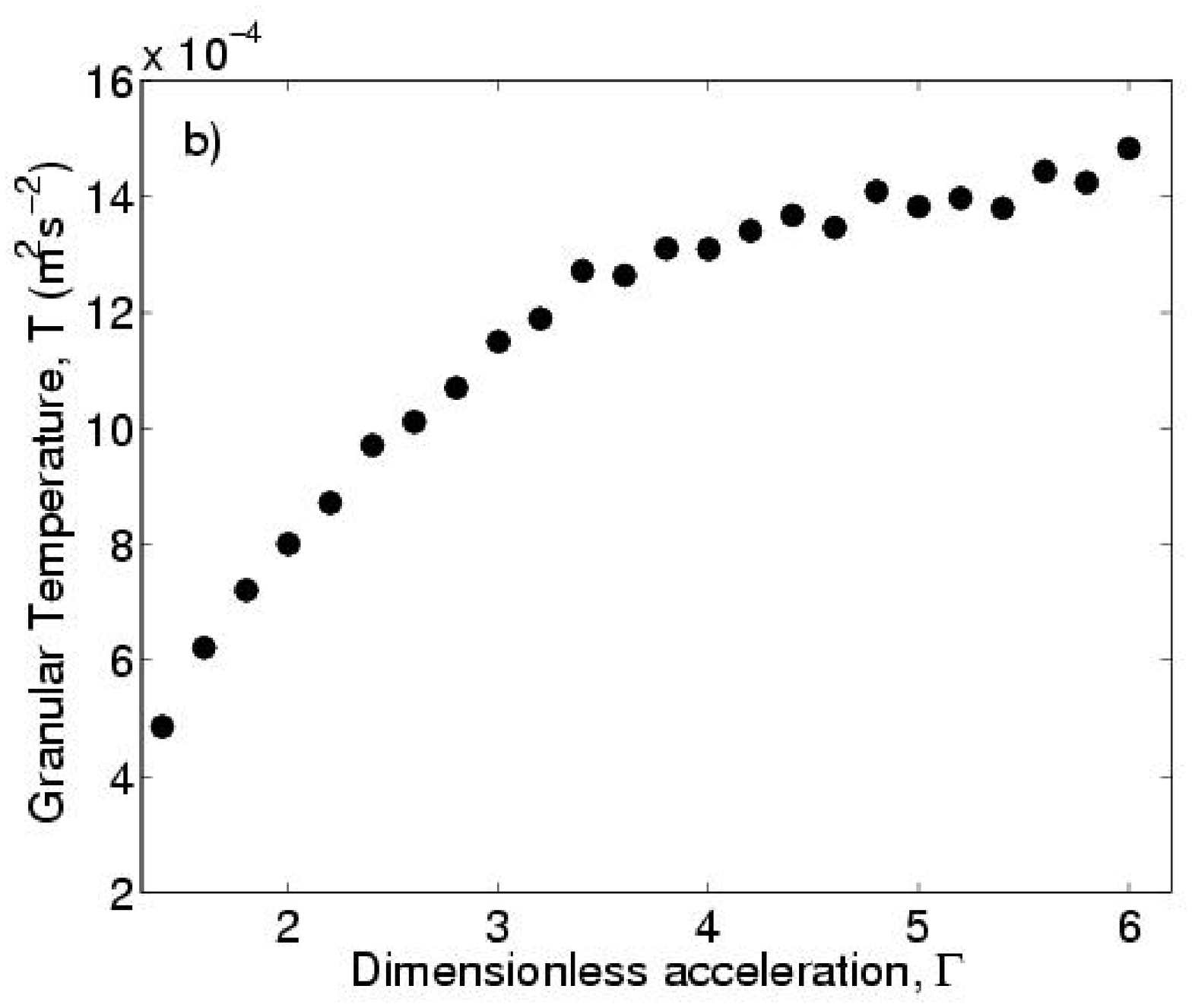}
        \caption{Dependence of the granular temperature, $T$, on driving parameters: (a) frequency, $f$ and (b) dimensionless acceleration $\Gamma$. The filling fraction is kept constant at $\phi=0.59$. \label{fig:temperature_forcing}}
    \end{center}
\end{figure}

For the case of using a flat bottom plate, the non-monotonic dependence
of $T$ on $\phi$ is more dramatic and it is difficult to attain
homogeneous states below $\phi< 0.4$. This is due to the fact that
at those filling fractions (the limiting case being the single
particle investigated in the previous section) the particles
perform small excursions and interact with their neighbors only
sporadically. Hence, little momentum is transferred onto the
horizontal plane. For $\phi>0.5$, there is an increase of
temperature with increasing filling fraction as interaction
between neighbors becomes increasingly more common up to $\phi\sim
0.7$ at which the curve for the flat bottom plate coincides with
the curve for the rough bottom plate. It is interesting to note
that the value at which this matching occurs is close to the point
of crystallization of disks in 2D, $\phi_c=0.716$
\cite{alder:1962,mitus:1997,reis:2006}. At this point, the large
number of collisions between neighboring particles thermalizes the
particles, independently of the details of the \emph{heating},
i.e. of whether a flat or rough plate is used.

The behavior of the layer was therefore more robust by using a rough bottom plate. Moreover, this allowed for a wider range of filling fractions to be explored, in particular in the low $\phi$ limit. Further detailed evidence for the advantage of using the rough bottom plate over a flat one is given in Appendix \ref{sec:appendix}. Hence, all results presented in the remainder of this paper correspond to experiments for which a rough glass plate was used for the bottom plate of the experimental cell.

\begin{figure}[t]
    \begin{center}
      \includegraphics[width=7.5cm]{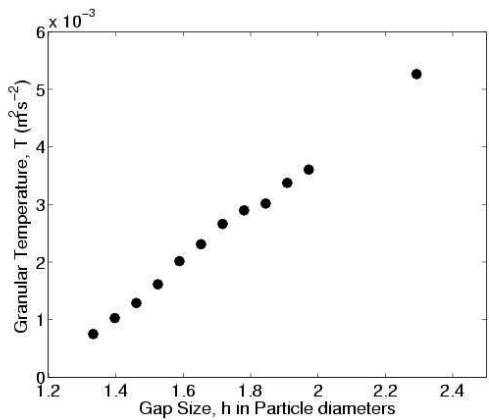}\\
        \caption{Dependence of the granular temperature, $T$, on cell gap, $h$, which was varied by changing the thickness of the inter-plate annulus. Filling fraction was kept constant at $\phi$=0.59 and the dimensionless acceleration $\Gamma$ was fixed to 4.
        \label{fig:temperature_cellgap}}
    \end{center}
\end{figure}

In Fig. \ref{fig:temperature_forcing} we present the dependence of
the granular temperature of the layer on the forcing parameters
for a fixed value of filling fraction, $\phi=0.59$. There is a
monotonic increase of $T$ with both $f$ and $\Gamma$.

The final control parameter that we investigated was the height of
the experimental cell which we varied by changing the thickness of
the inter-plate annulus by using precision spacers. The range
explored was $1.3D<h<2.3D$. Note that for $h=D$ there would be no
clearance between the spheres and the glass plates and one would
therefore expect for no energy to be injected into the system. On
the other hand for $h>2D$, spheres could overlap over leach other
and the system ceases to be quasi-2D. The dependence of
the granular temperature on cell gap is plotted in Fig.
\ref{fig:temperature_cellgap}. It is interesting to note that $T$
appears to depend linearly on the gap height.

One can therefore regard changing $h$, $f$ and $\Gamma$ as a way
of varying the temperature of the granular fluid, which can this
way be tuned up to a factor of eight.


\section{Probability density functions of velocities}
\label{sec:pofv}

\begin{figure}[b]
    \begin{center}
            \includegraphics[width=0.82\columnwidth]{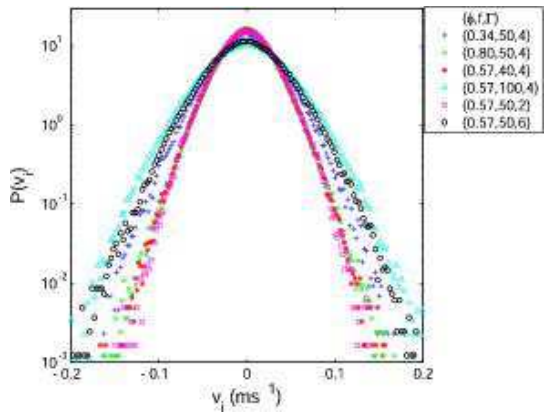}
                   \caption{(a) Probability distribution function of velocities (PDF),
                   $P(v)$, for specific values of filling fraction ($\phi$), frequency ($f$) and dimensionless acceleration ($\Gamma$).
                   See legend for specific values.\label{fig:velocitydistributions_mpers}}
    \end{center}
\end{figure}

\begin{figure}[b]
    \begin{center}
            \includegraphics[width=0.82\columnwidth]{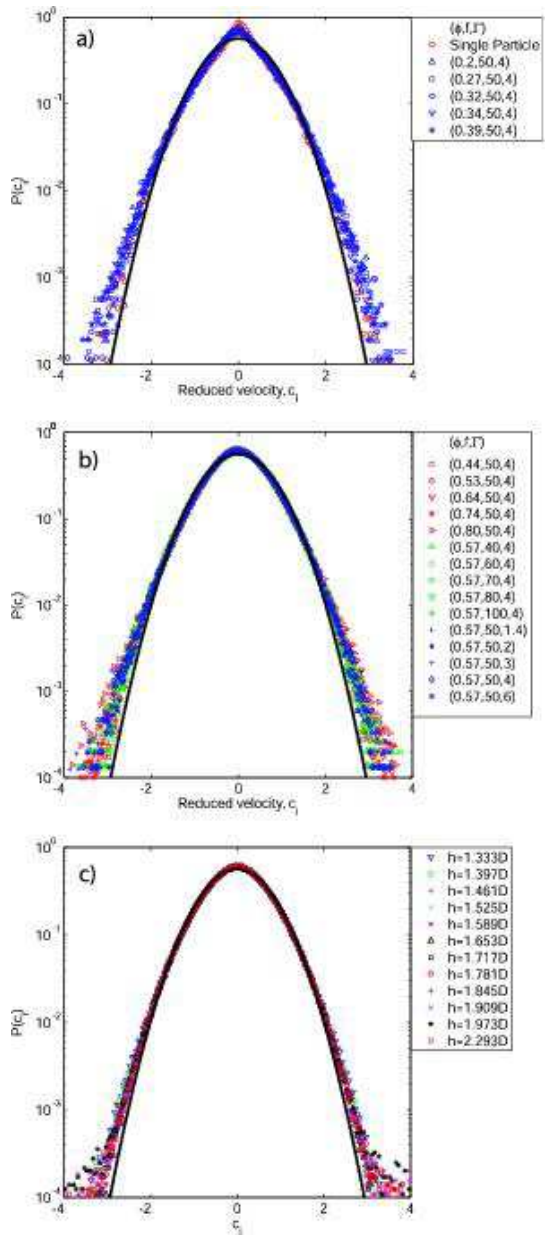}
                   \caption{(a)-(b)PDF of velocities,
                   $P(c)$, in which all velocities for each distribution were
                   normalized by its standard deviation. Data for a range of $\phi$,
                   $f$ and $\Gamma$. (c) PDF of velocities for range of values of cell gap,
                   $h$. Other parameters set to $\phi=0.59$, $f=50Hz$ and $\Gamma=4.0$.
                   (a-c) See legend for specific values of the experimental parameters. The solid lines are Gaussian with unit standard deviation.\label{fig:velocitydistributions}}
    \end{center}
\end{figure}

We now turn to the distribution functions of particle velocities
under various conditions of filling fraction, frequency,
dimensionless acceleration and gap height. In Fig.
\ref{fig:velocitydistributions_mpers} we plot the
probability density function of a velocity component, $P(v_i)$, where the index $i$ represents the component $x$ or $y$, for specific values of
$\phi$, $f$ and $\Gamma$. From now on we drop the index in $v_i$ since we showed in the previous section that the dynamics of the system is isotropic in $x$ and $y$. Moreover, when we refer to \emph{velocity} we shall mean \emph{velocity component, $i$}. The width of the observed $P(v)$ differs
for various values of the parameters. This is analogous to the fact that the
temperature (i.e. the variance of the distribution) is different for various states set by $\phi$,
$f$ and $\Gamma$, as discussed in the previous Sec.. It is
remarkable, however, that all the $P(v)$ distributions can be
collapsed if the velocities are normalized by the characteristic
velocity,
\begin{equation}
     v_o=\sqrt{2\langle v^2\rangle}=\sqrt{2T}
     \label{eqn:characteristic_vo}
\end{equation}
where the brackets $\langle . \rangle$ represent time and ensemble
averaging over all the particles in the field of view of the
imaging window. This collapse is shown in Figs.
\ref{fig:velocitydistributions}(a) and (b) and was accomplished not only for various
filling fractions  but also for a range of
frequencies  and dimensionless accelerations. 

To highlight the quality of the collapse we have separately plotted
$P(c)$ for the four lowest values of filling fraction ($\phi={0.27,
0.32, 0.34, 0.39}$) along with the $P(c)$ for a single particle in
Fig. \ref{fig:velocitydistributions}(a) and all other data in the
ranges $0.44<\phi<0.8$, $40 Hz<f<100 Hz$ and $1.4 < \Gamma < 6$ in
Fig. \ref{fig:velocitydistributions}(b). At low values of the filling
fraction ($\phi<0.44$) the collapse is satisfactory but deviations are
seen at low $c$. In particular, near $c=0$ the distributions exhibit a
sharp peak with a clearly discontinuous first derivative which
reflects the fact that for these value of low $\phi$ the resulting
gases are not collisionally driven but are dominated instead by the
underlying thermostat (c.f. distribution for single
particle). However, this sharp peak becomes increasingly smoother as
the filling fraction is increased, presumably due to the increasing
number of particle collisions, and by $\phi\sim0.44$, it has
practically disappeared.  On the other hand, for $\phi>0.44$ the
collapse onto a universal curve is remarkable for such a wide range of
$\phi$, $f$ and $\Gamma$. The choice of the lower bounds for frequency
and acceleration, $f=40Hz$ and $\Gamma=1.4$, in the data plotted will
be addressed in Sec. \ref{sec:pofv_lowv}. Moreover, this collapse
of $P(v)$ is also attained for various values of the gap height
($1.3D<h<2.3D$), as shown in
Fig. \ref{fig:velocitydistributions}(c). From now onwards, we shall
perform our analysis in terms of the reduced velocity, $c=v/v_o$.

We stress the universal collapse of the experimental velocity
distribution functions for a wide range of parameters and the clear
deviation from the Maxwell-Boltzmann (Gaussian) distributions of
Eqn. (\ref{eqn:MB}) -- solid curve in Fig.
\ref{fig:velocitydistributions}(b) and (c). This is particularly
visible at high velocities where there is a significant overpopulation
of the distribution tails in agreement with previous theoretical
\cite{vannoije:1998}, numerical \cite{moon:2001} and other
experimental work \cite{rouyer:2000,aranson:2002,zon:2004}. To
quantify these deviations from Gaussian behavior, our analysis will be
twofold.  First we shall analyze the non-Gaussian tails of the velocity
distributions (Sec. \ref{sec:pofv_tails}). Even though the
tails correspond to events associated with large velocity the
probability of them happening is extremely low. However, this overpopulation at the tails implies that, if the distribution is to remain normalized with a standard deviation given by the average kinetic energy per particle, clear deviations from Gaussian at the central regions of low $c$ (high probability) of the distribution must necessarily also be observed. In Sec. \ref{sec:pofv_lowv} we study this deviations in the context of Sonine corrections introduced in Sec. \ref{sec:kinetic_theory}.

In Appendix \ref{sec:appendix} we provide the data corresponding to Fig. \ref{fig:velocitydistributions_mpers} and Figs. \ref{fig:velocitydistributions}(a,b) of the velocity distributions obtained using the optically flat bottom plate, instead of the rough plate used here. We shall show that the $P(v)$s for the flat bottom plate exhibit  much stronger non-universal deviations from Gaussian behavior and that the collapse of $P(c)$ is highly unsatisfactory. This highlights, as mentioned in Sec. \ref{sec:granaulartemperature}, the considerable advantage in our experimental technique of using the rough plate to generate the granular fluid.

        \subsection{Deviation at large velocities -- The tails}
            \label{sec:pofv_tails}

Recently, van Noije and Ernst \cite{vannoije:1998} have made the
theoretical prediction that the high energy tails of the velocity
distributions of a granular gas \emph{heated} by a stochastic
thermostat should scale as stretched exponentials of the form,
     \begin{equation}
     P(c)~\sim \exp\left(-A c^{3/2}\right),
          \label{eqn:vannoijeprediction}
     \end{equation}
where $A$ is a fitting parameter. Note that the theoretical
argument of van Noije and Ernst involves a high $c$ limit and
hence one does not expect this stretched exponential form to be
valid across the whole range of the distribution.

\begin{figure}[t]
    \begin{center}
            \includegraphics[width=0.8\columnwidth]{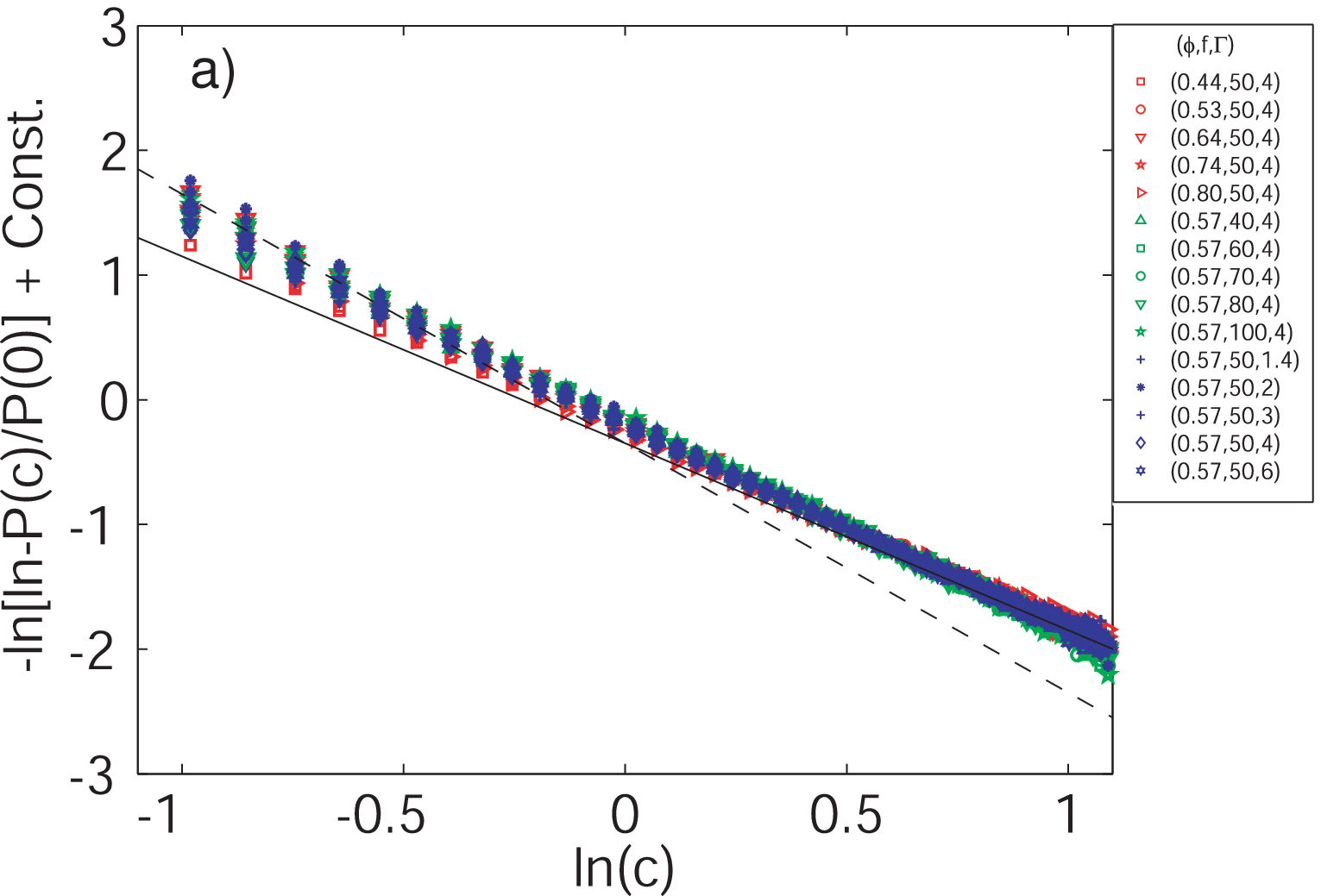}
            \includegraphics[width=0.8\columnwidth]{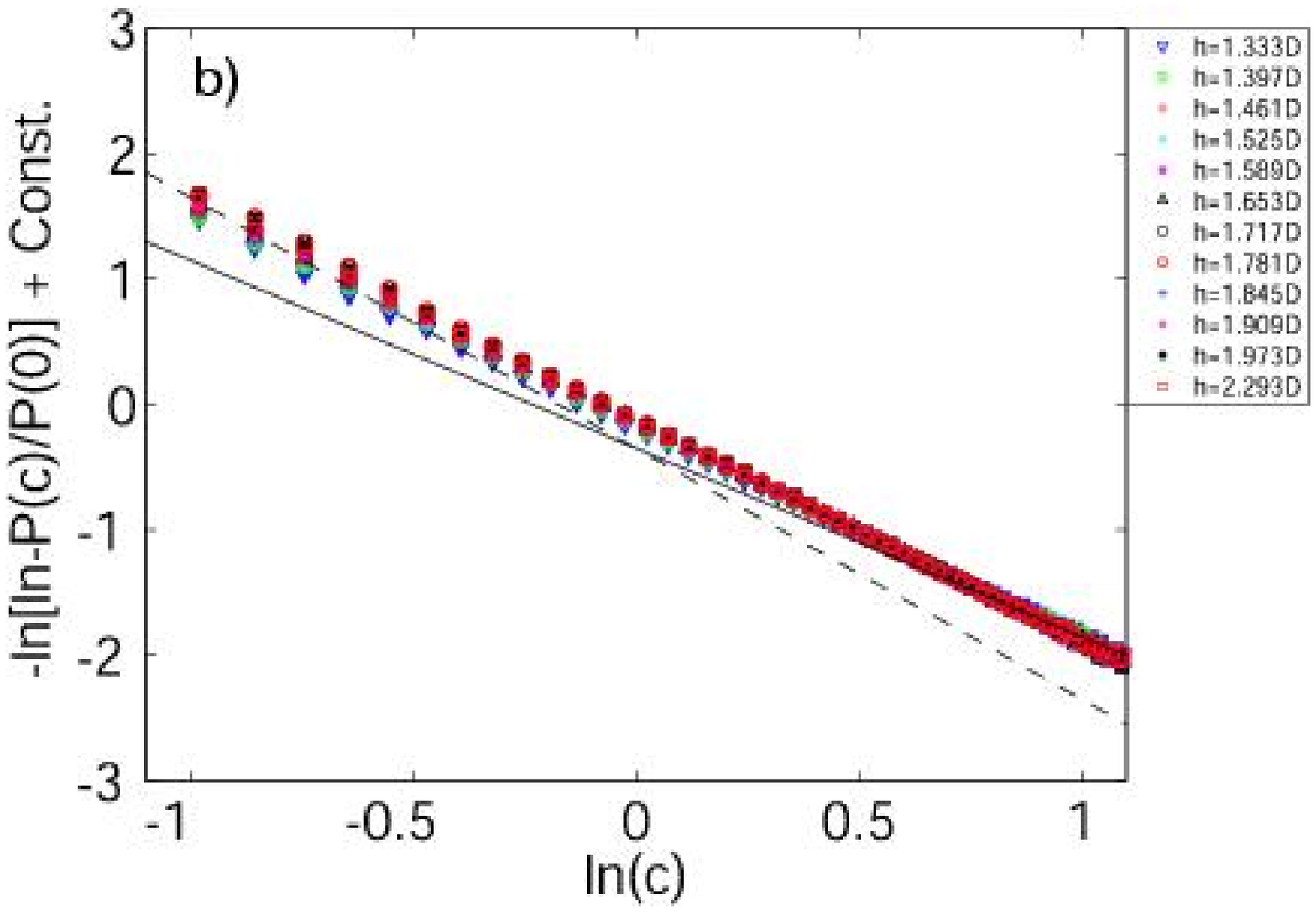}
        \caption{(a) Tails of  $P(c)$ for specific values of filling fraction ($\phi$), frequency ($f$) and dimensionless acceleration ($\Gamma$) and (b) gap height ($h$).The solid lines correspond to stretched exponentials
        of the form $\sim \exp(-Ac^{{3/2}})$ whereas the dashed lines correspond to the
        Gaussian behavior of the form $\sim \exp(-Ac^{{2}})$. \label{fig:tails_phi_freq_gamma}}
    \end{center}
\end{figure}

To check the applicability of Eqn. (\ref{eqn:vannoijeprediction}) to describe our data, we have calculated the quantity $q(c)=-\ln\left( -\ln (P(c))\right)$, which we plot in Fig. \ref{fig:tails_phi_freq_gamma}(a) for a variety of filling fractions, frequencies and dimensionless accelerations. Indeed, within the ranges considered and, in the limit of large $c$, $q(c)$ tends to a straight line with slope -3/2 for all values of the control parameters, in excellent agreement with the scaling of Eqn. (\ref{eqn:vannoijeprediction}). In Fig. \ref {fig:tails_phi_freq_gamma}(a) we have excluded filling fractions $\phi<0.44$, since as discussed in the previous Sec., the gases obtained in this range are dominated by the thermostat rather than collisions and the main ingredient for the prediction of Eqn. (\ref{eqn:vannoijeprediction}) is the role of inelastic collisions. This behavior of the tails of the velocity distribution is also observed for various values of the gap height, $h$, of the experimental cell, as shows in Fig. \ref{fig:tails_phi_freq_gamma}(b). Note the crossover from Gaussian-like (dashed line with slope -2) to the stretched exponential (solid line with slope -3/2)  at $c\sim 1$, which is particularly clear at the larger values of $\phi$. For low $\phi$ the tails of the distribution tend to be closer to the stretched exponential throughout the range $-1.1<\log(c)<1.1$. We stress that Eqn. (\ref{eqn:vannoijeprediction}) has been derived in the limit of large $c$ and is therefore only expected to be valid at the large $c$ end of the tails. A. Barrat \emph{el. al.} \cite{barrat:2003,barrat:2005} have investigated the range of validity in a numerical model with a stochastic forcing and found that the stretched exponential scaling of the tails should only be valid for very high velocities corresponding to probabilities lower than $10^{-6}P(0)$. The fact that we find this scaling at considerable lower velocities, i.e. in regions of higher probability, (in agreement with other experimental work \cite{rouyer:2000}) raises some questions regarding the detailed comparison between theory and experiments.

        \subsection{Sonine corrections to the distribution}
            \label{sec:pofv_lowv}

Having looked at the tails of the distributions, i.e. at
large $c$, we now extend our deviation analysis  based on an expansion method to the full $c$ range of the distributions. This highlights in particular deviations from Gaussian behavior in the central high probability regions of $P(c)$ near $c=0$. In Sec. \ref{sec:kinetic_theory} we discussed that in the
solution of the Enskog-Boltzmann equation for inelastic particles
in a stochastic thermostat, a Sonine expansion is usually
performed such that the deviations from Gaussian are described by
a Sonine Polynomial (i.e. a 4th order polynomial with well defined
coefficients) multiplied by a numerical coefficient $a_2$.

\begin{figure}[b]
    \begin{center}
            \includegraphics[width=\columnwidth]{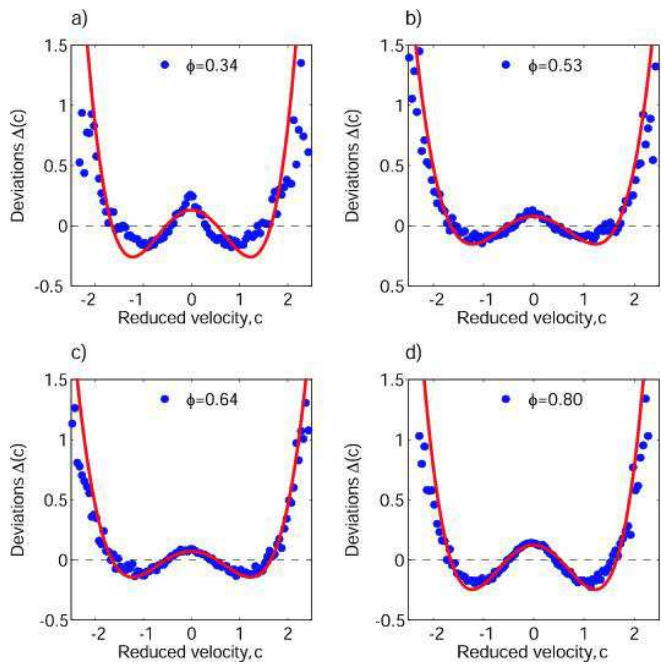}
        \caption{Experimental deviation function from Gaussian behavior, $\Delta(c)$ for
        (a) $\phi=0.34$, (b) $\phi=0.53$, (c) $\phi=0.64$ and (d) $\phi=0.80$.
        The solid line is the order-two Sonine polynomial of the
        form
        $a_{2}(1/2c^{4}-3/2c^{2}+3/8)$ where $a_2$, the second Sonine coefficient, is the only adjustable parmeter . \label{fig:order4deviations}}
    \end{center}
\end{figure}

\begin{figure}[t]
    \begin{center}
            \includegraphics[width=\columnwidth]{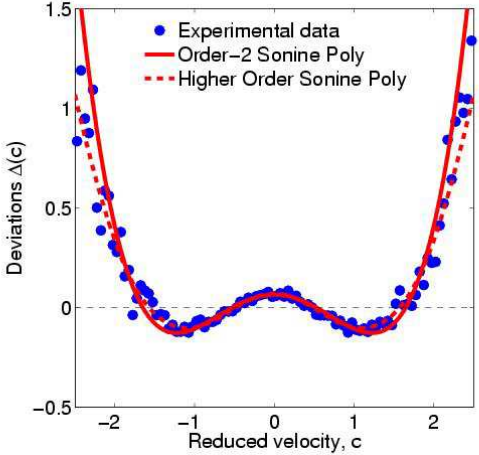}
        \caption{Experimental deviation function from Gaussian behavior,
        $\Delta(c)$ for $\phi=0.66$. The solid line is the Sonine polynomial
        $a_{2}(1/2c^{4}-3/2c^{2}+3/8)$ with one single fitting parameter:
        $a_{2}=0.171$. The dashed line is the higher order Sonine
        polynomial description of the form $\sum_{p=2}^6 a_p S_p(c^2)$
        with the following (five fitting parameters) Sonine coefficients; $a_2=0.1578$, $a_3=-0.0656$, $a_4=0.1934$,
        $a_5=-0.1637$ and $a_6=0.0832$. \label{fig:deviation_phi_0_66}}
    \end{center}
\end{figure}

\begin{figure}[t]
    \begin{center}
            \includegraphics[width=\columnwidth]{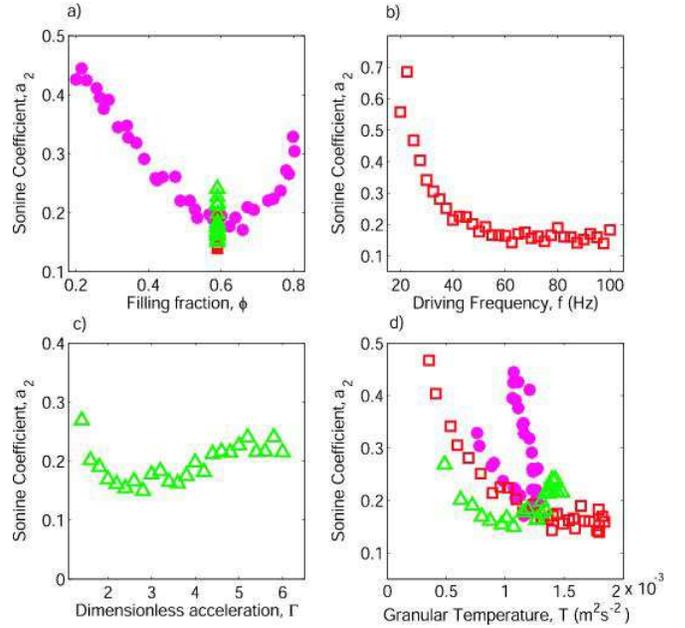}
        \caption{Experimentally determined order-two Sonine coefficient $a_2$ as a function
        of (a) filling fraction ({\large{$\bullet$}}), (b) driving frequency  ($\square$), (c) dimensionless acceleration ($\triangle$) and (d) granular temperature. The filling fraction is
        kept constant at $\phi=0.59$ while exploring the dependence of $a_2$ on $f$ and $\Gamma$.
        \label{fig:sonine_coeff}}
    \end{center}
\end{figure}

To check the validity of this assumption of the Kinetic Theory, we
shall follow a procedure analogous to that employed in the
numerical study of Ref. \cite{moon:2001}. We calculate the
deviation, $\Delta(c)$, of the experimental velocity
distributions, $P(c)$, from the equilibrium Maxwell-Boltzmann, $f_{MB}$, such that,
     \begin{equation}
          P(c)=f_{MB} \left( 1+ \Delta(c) \right),
          \label{eqn:deviations}
     \end{equation}
where $f_{MB}$ is given by Eqn. (\ref{eqn:MB}). By comparing Eqn.
(\ref{eqn:deviations}) to Eqn.
(\ref{eqn:theory_truncation}) for the theoretical case of
inelastic particles under a stochastic thermostat, we expect that
	\begin{equation}
		P(c)=f_{MB}(1+a_2S_2(c^2)),
		\label{eqn:corrected:distributions}
	\end{equation}
such that the experimental deviations from equilibrium take the form of the Sonine polynomial of order-two, $S_2(c^2)$, i.e.
     \begin{equation}
          \Delta(c)=a_2\left( 1/2c^4 - 3/2c^2 + 3/8 \right),
          \label{eqn:sonine_order2}
     \end{equation}
where the parameter $a_2$ can be directly related to the kurtosis $K$
of the distribution as $a_{2}=K/3-1$. However, because of sampling
noise at high $c$, we have taken $a_{2}$ as a fitting parameter rather
than determining it directly from $K$, although in most cases there is
little difference. Note that the fitting of the experimental data to
Eqn. (\ref{eqn:corrected:distributions}) was done over the whole range
of the distribution with a relative weighting given by the
corresponding probability.  In Fig. \ref{fig:order4deviations}(a--d)
we plot these experimental deviation from equilibrium, $\Delta(c)$,
for four representative values of filling fraction, along with the
order-2 Sonine polynomial (solid curve) as given by Eqn.
(\ref{eqn:sonine_order2}) and fitting for $a_2$. At low filling
fractions, for example $\phi=0.34$ --
Fig. \ref{fig:order4deviations}(a) -- the second order Sonine fit is
approximate but unsatisfactory for small $c$ where a sharp cusped peak
is present. This cusp is reminiscent of the underlying dynamics of the
rough plate thermal bath as evidenced by the cusp at $c=0$ for the
single particle distribution presented in
Fig. \ref{fig:pofv_singleparticles_pofv}(c) and discussed
previously. On the other hand, for larger values of the filling
fraction (in particular for $\phi>0.44$) the experimental data is
accurately described by the second order Sonine polynomial.

\begin{figure}[t]
    \begin{center}
            \includegraphics[width=\columnwidth]{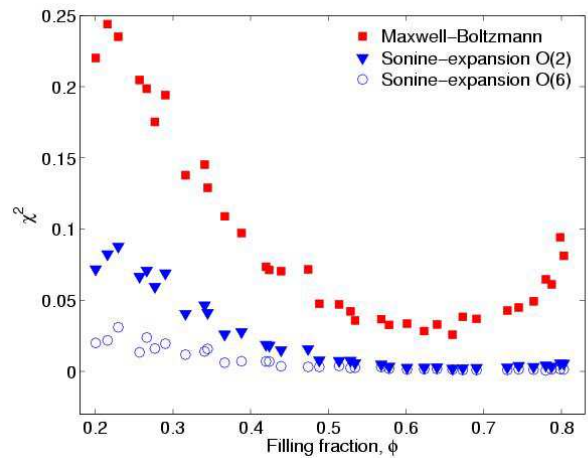}
        \caption{Deviations (quantified by $\chi^2$), as a function of filling fraction, of the experimental velocity distributions from Maxwell-Boltzmann distribution ($\square$), from the velocity distribution function with a order-two Sonine polynomial correction ($\triangledown$) and from the velocity distribution function with a order-six Sonine polynomial expansion correction ({\large{$\circ$}}) .          \label{fig:deviations}}
    \end{center}
\end{figure}

To further evaluate the relevance of the Sonine polynomials to
describe the experimental deviation from Maxwell-Boltzmann, in
Fig. \ref{fig:deviation_phi_0_66} we plot the experimental $\Delta(c)$
along with both the order-two ($p=2$) Sonine polynomial and the higher
order Sonine expansion of the form $\sum_{p=2}^6 a_p S_p(c^2)$. The
second order Sonine polynomial alone is responsible for the largest
gain in accuracy of the corrections from MB (horizontal dashed
line). The higher order Sonine expansion indeed provide a better fit
but the overall improvement is only marginal (see
Fig. \ref{fig:deviations}.)

In Fig. \ref{fig:sonine_coeff}(a) we present the dependence of the
second Sonine coefficient $a_2$, the single fitting parameter in
Eqn. (\ref{eqn:sonine_order2}), as a function of filling fraction.
The coefficient $a_2$ initially decreases with increasing filling
fraction up to $\phi=0.65$, after which it shows a rapid rise. We also
study the dependence of $a_2$ on the driving frequency ($f$) and
dimensionless acceleration ($\Gamma$) for a single value of the
filling fraction $\phi=0.59$, which we present in
Figs. \ref{fig:sonine_coeff}(b) and (c), respectively. For lower
values of frequency (20-40Hz) we get a monotonic drop in $a_2$ with
increasing frequency. For $f>40Hz$, the coefficient $a_2$ then levels
off and remains approximately constant at $a_2=0.171\pm0.023$. For
$\Gamma>2.0$, the coefficient $a_2$ remains approximately constant at
$a_2=0.193\pm0.029$. We have plotted these two datasets at fixed
$\phi$ and varying $f$ and $\Gamma$ back in
Fig. \ref{fig:sonine_coeff}(a) and the behavior of $a_2$ is found to
be consistent with the previous dataset with varying $\phi$ at fixed
$(f,\Gamma)=(50Hz,4.0)$. It is interesting to note that a scatter of
points is obtained if $a_2$ for the three previous datasets is plotted
as a function of the corresponding granular temperatures. Moreover,
for the data-set for varying $\phi$ the dependence is clearly not
single valued, i.e. one can find two distinct values of $a_2$ for a
single value of temperature.  All this suggests that the temperature
does not set $a_2$. Instead, $a_2$ is a strong function of $\phi$ and
is only weakly dependent on $f$ and $\Gamma$, provided that $f>40Hz$
and $\Gamma>2$. This is in agreement with inelastic hard-sphere
behavior where the only two relevant parameters are thought to be the
filling fraction and the coefficient of restitution (not investigated
in the present study).

\begin{figure}[t]
    \begin{center}
            \includegraphics[width=0.95\columnwidth]{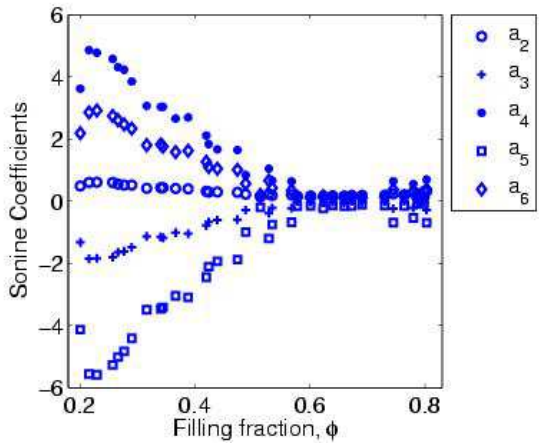}
        \caption{Dependence of higher order coefficients of the Sonine expansion ($a_2$ to $a_6$) on filling fraction.
            \label{fig:Sonine_higherorder}}
    \end{center}
\end{figure}

Next, in Fig. \ref{fig:Sonine_higherorder} we present the dependence on filling fraction of the experimentally determined five non-zero Sonine coefficients in the sixth order expansion of $\Delta(c)\sim\sum_{p=2}^6 a_p S_p(c^2)$. In the theoretical analysis,  the higher order terms (order 3 and
above) in the Sonine polynomial expansion are typically neglected
to simplify the calculations, the claim being that this results in
no significant loss of accuracy. In our experimentally generated fluids the higher order Sonine coefficients assume values less than 1 for  filling fractions above $\phi = 0.5$. From this it is clear that it is unnecessary to consider orders higher than two for intermediate and high filling fractions.  For $\phi<0.5$, however, they are significant and the finite values of $a_3$,..., $a_6$ are required to fit the increasingly larger regions of stretching exponential tails which become progressively more wider and propagate towards  smaller velocities. As we mentioned in the discussion of Fig. \ref{fig:tails_phi_freq_gamma}(a) in the previous Sec., when the filling fraction decreases, the dynamics of the layer starts resembling that of the underlying thermal bath set by the rough plate. There the distribution,  at low $c$ becomes increasingly more like a stretched exponential throughout the full $c$ range, rather than only in its tails as is the case at larger values of $\phi$. This finding is analogous to a result from Brilliantov and P\"oschel \cite{brilliantov:2006} who found both theoretically and numerically that the magnitude of the higher-order Sonine coefficients can grow due to the an increasing impact of the overpopulated high-energy tails of the distribution function, just like in our system at low $\phi$. Their analysis was however performed as a function of the coefficient of restitution, $\epsilon$, rather than filling fraction. In their study a single second order Sonine polynomial becomes insufficient to represent the deviation from Gaussian at low values of $\epsilon$. Moreover, they worked on the Homogeneous Cooling State (HCS), but they suggest that their results should also apply to granular gases with a thermostat, like in our case. A more direct comparison with this theoretical analysis is open to further investigation. 

We stress that a finite value of the higher order coefficients does
not necessarily imply a large correction to the deviations
$\Delta(c)$. To explore further this point we now quantify the
deviations of our experimental velocity distribution from all the
three models we have considered: 1) the equilibrium Maxwell-Boltzmann
distribution, 2) the velocity distribution function with order-2
Sonine polynomial expansion and, 3) the velocity distribution function
with higher order Sonine polynomial terms. In
Fig. \ref{fig:deviations} we plot these deviation of experiments from
the models as quantified by $\chi^2=\sum_{i=1}^N
(P(c_i)_{exp}-P(c_i)_{model})^2$, for the full range of filling
fractions. We clearly see significant deviation of the experimental
data from Maxwell-Boltzmann distribution for all $\phi$, whereas the
velocity distributions with the order-2 Sonine polynomial term shows a
considerably better agreement with the experimental data across the
whole range of $\phi$. Even though the higher order Sonine polynomial
expansion characterizes the experimental data more closely, these
higher order contributions are modest compared with the large gain in
accuracy from the order-2 Sonine polynomial alone, and in the range of
$\phi>0.44$ provide no significant improvement. Thus the order-2
description obtained by neglecting the higher order Sonine terms is a
reasonably good approximation to describe the experimental data.

\section{Conclusion}
    \label{sec:conclusion}

In conclusion, we have developed an experimental model system for a quasi-2D granular layer, under homogeneous stochastic driving. Our experimental technique has allowed us to randomly thermalize a granular fluid over a wide range of filling fractions. Our study was centered on the dynamics of the system, in particular the statistics of velocities. The temperature of the experimental granular fluid could be adjusted by varying the system's control parameters, namely the filling fraction, the frequency and acceleration of the driving and the gap height, with no significant change in the nature of the velocity probability distribution functions. We have found an excellent collapse of the distribution functions if the particle velocities are scaled by a characteristic velocity $v_o$  (the standard deviation of the distributions).

However, the obtained distribution are non-Gaussian. We have analyzed the deviations from Gaussian behavior in two distinct regimes. Firstly, we looked at the shape of the distribution tails which scaled as stretched exponentials with exponent $-3/2$. Secondly, we have performed an expansion method that highlights the deviations from a Gaussian  at low velocities  (near $c=0$) and found them to be Sonine-like, i.e. polynomial of order four with fixed coefficients. This way we have determined the validity of some important assumptions in the Kinetic Theory of randomly driven fluids. It is surprising that this formalism seems applicable at filling fractions as high as $\phi=0.80$, whereas Kinetic Theory is usually thought to breakdown at much lower values of $\phi$. Finally, we have looked at Sonine polynomial expansion with higher order terms and concluded that it is sufficient to retain only the leading order (first non-vanishing) term  to maintain reasonable accuracy of the polynomial expansion.  Therefore, we can accurately characterize the single particle velocity distribution function by introducing a single extra coefficient $a_2$, in addition to the 0th, 1st and  2nd moments used for fluids at equilibrium. The coefficient $a_2$ has universal character in the sense that it is  a strong function of filling fraction and only depends weakly on the other experimental parameters such as the driving frequency and acceleration. 

Having determined the base state of our randomly driven granular fluid, we hope that an experimental system such as ours can be used further as a laboratory test bed of  some basic assumptions of Kinetic Theory. In future work, it would also be of interest to investigate whether particle-particle velocity correlations \cite{prevost:2002} are present in our system but this is beyond the scope of the current investigation. 

To our knowledge, this is the first time that the Sonine corrections of the central high probability regions of the velocity distributions have been measured in an experimental system, in agreement with analytical predictions. This should open way to further theoretical developments, crucial if we are to develop much desired predictive models for granular flows with practical relevance.

\appendix

\section{Probability density functions of velocities using the bottom flat plate}
\label{sec:appendix}

\begin{figure}[b]
    \begin{center}
           \includegraphics[width=0.82\columnwidth]{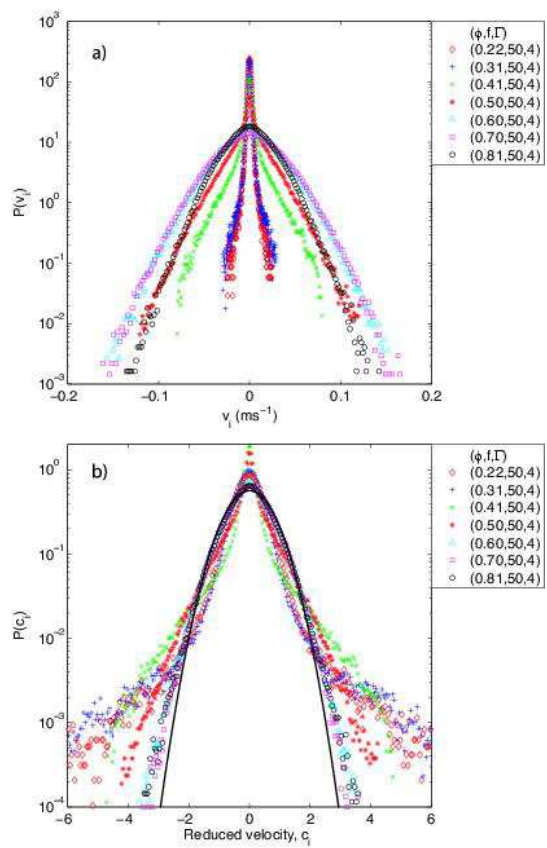}         
                    \caption{ PDF of velocities for an experimental cell where both top and bottom plates were optically flat. (a) $P(v)$ (b) $P(c)$, in which all velocities for each distribution were normalized by its standard deviation. The solid line is a Maxwell-Boltzmann distribution with unit standard deviation. Data for a range of parameters: $\phi$, $f$ and $\Gamma$  (see legends for specific values). \label{fig:velocitydistributions_flat}}
    \end{center}
\end{figure}

In this Appendix we discuss the nature of the velocity distributions obtained using a flat bottom plate as in the system of Olafsen and Urbach \cite{olafsen:1999}. The distributions $P(v)$  are presented in Figs. \ref{fig:velocitydistributions_flat}(a)  for a wide range of filling fractions. Unlike the case of using a rough plate, here there is an enormous  variation in the shape of $P(v)$. At low $\phi$  the distributions are sharply peaked near $c=0$. The value of $P(v=0)$ can differ by over a factor of 20 between low and high $\phi$. In Fig. \ref{fig:velocitydistributions_flat}(b) the distributions $P(c)$ -- obtained after rescaling all velocities by the characteristic velocity $v_o$ given by Eqn. (\ref{eqn:characteristic_vo}) --  do not collapse onto a single curve and show considerably larger deviations from a Maxwell-Boltzmann (solid line). This distributions only attain the general form found for the rough case at filling fractions larger than  $\phi>0.6$ and this is in agreement with the point at which the granular temperature of both cases coincide -- Fig. \ref{fig:temperatures_flat_rough}. For this high values of $\phi$, both in the rough and flat cases, the contribution to the thermalization is more important from the large number inter-particle collisions rather than the underlying thermostat. Hence we expect the analysis we have performed throughout this paper to be applicable to the flat case, for $\phi>0.6$, but not below this point. 

In addition to the discussion on the granular temperature of Sec. \ref{sec:granaulartemperature}, these results present further evidence for the significant advantage of using a rough bottom plate to generate the thermostat to \emph{heat} the granular fluid in a way that may be directly compared to theoretical work on inelastic hard-spheres driven by a stochastic thermostat.

This work is funded by The National Science Foundation, Math, Physical
Sciences Department of Materials Research under the Faculty Early
Career Development (CAREER) Program (DMR-0134837). PMR was partially
funded by the Portuguese Ministry of Science and Technology under the
POCTI program and the MECHPLANT NEST-Adventure program of the European
Union.

\bibliography{velocity_statistics,newsand}
\bibliographystyle{apsrev}

\end{document}